\documentclass[11pt]{article}

\usepackage{jheppub}
\usepackage{xcolor,comment}
\usepackage{url}
\usepackage{hyperref}
\usepackage{combelow}
\definecolor{purple}{rgb}{0.8,0,0.6}

\newcommand{\eq}[1]{(\ref{#1})}

\hypersetup{
  colorlinks,
  citecolor=blue,
  linkcolor=blue,
  urlcolor=blue}
\usepackage{amsmath}
\usepackage{amsfonts}
\usepackage{amssymb}
\usepackage{color}
\usepackage{tikz}
\usepackage{feynmf}
\usepackage{bm}
\usepackage{cleveref}
\usepackage{slashed}
\usepackage{bm}
\usepackage{ dsfont }
\newcommand{\beq}{\begin{equation}}
\newcommand{\eeq}{\end{equation}}
\newcommand{\beqa}{\begin{eqnarray}}
\newcommand{\eeqa}{\end{eqnarray}}
\newcommand{\beqn}{\begin{eqnarray}}
\newcommand{\eeqn}{\end{eqnarray}}
\newcommand{\bs}{\boldsymbol}
\newcommand{\cL}{{\cal L}}

\newcommand{\avr}[1]{{\left\langle #1 \right\rangle}}
\usepackage{graphicx,amssymb,amsmath,subfigure}
\usepackage{xcolor}

\usepackage{graphicx}

\newcommand{\dd}{\mathrm{d}}

\begin{document}

\title{A new  scale anomaly in Dirac matter}

\author[a,b,c]{Matteo Baggioli}
\emailAdd{b.matteo@sjtu.edu.cn}
\author[d,e]{, Maxim N. Chernodub}
\emailAdd{maxim.chernodub@univ-tours.fr}
\author[f]{, Karl Landsteiner}
\emailAdd{karl.landsteiner@csic.es}
\author[g]{, Alessandro Principi}
\emailAdd{alessandro.principi@manchester.ac.uk}
\author[h,i]{, Mar\'ia A. H. Vozmediano}
\emailAdd{mahvozmediano@gmail.com}
\affiliation[a]{School of Physics and Astronomy, Shanghai Jiao Tong University, Shanghai 200240, China }
\affiliation[b]{Wilczek Quantum Center, School of Physics and Astronomy, Shanghai Jiao Tong University, Shang-hai 200240, China}
\affiliation[c]{Shanghai Research Center for Quantum Sciences, Shanghai 201315, China}
\affiliation[d]{Institut Denis Poisson UMR 7013, Universit\'e de Tours, 37200 France}
\affiliation[e]{Department of Physics, West University of Timi\cb{s}oara,  Bd.~Vasile P\^arvan 4, Timi\cb{s}oara 300223, Romania}
\affiliation[f]{Instituto de F\'isica Te\'orica UAM/CSIC, 
Nicol\'as Cabrera 13-15, Cantoblanco, 28049 Madrid, Spain }
\affiliation[g]{Department of Physics and Astronomy, University of Manchester,
 Oxford Road, Manchester M13 9PL, United Kingdom}
\affiliation[h]{Instituto de Ciencia de Materiales de Madrid, CSIC, Cantoblanco; 28049 Madrid, Spain.}
\affiliation[i]{Donostia International Physics Center, Paseo Manuel de Lardiz\'abal 4, 20018 San Sebasti\'an, Spain}

\abstract{The dynamics of Dirac semimetals are modeled at low energies by the massless Dirac Hamiltonian with the Fermi velocity replacing the velocity of light. The classical action is scale invariant. In 3D materials, Coulomb interactions induce a conformal anomaly associated to the charge renormalization already known in quantum field theory. In this work, we describe a new  conformal anomaly induced by the running of the Fermi velocity that applies to Dirac semimetals in two and three dimensions. The case of graphene is particularly interesting. We analyze the anomaly and explore its thermodynamic and  hydrodynamic consequences. The anomaly modifies the propagation speed of hydrodynamic sound waves, alters the thermodynamic equation of state, and induces a non-vanishing bulk viscosity proportional to the beta function of the Fermi velocity.}

\maketitle
\section{Introduction}

Conformal (or scale) invariance\footnote{Although there are subtle differences between conformal and scale invariance for systems lacking Lorentz invariance \cite{Nakayama:2013is}, we will use the two names indistinctly.} is a concept that permeates many branches of physics, from critical points in phase transitions to cosmology, relativity, particle physics, and condensed matter \cite{10.1093/acprof:oso/9780199577224.001.0001,doi:10.1098/rsif.2017.0662,Zinn-Justin:2010}. In general, any physical system with no dimensionful parameters (massless particles and dimensionless coupling constants) is scale invariant. The conservation of the associated Noether current implies the tracelessness of the energy-momentum tensor of the system $T^{\mu\nu}$,
\begin{equation}
    T^\mu_\mu=\partial^\mu J_\mu,
\end{equation}
up to the divergence of a virial current $J^\mu$ \cite{Nakayama:2013is}. \parshape=0

Under a Quantum Field Theory (QFT)  point of view, strictly renormalizable theories (dimensionless coupling constants) with massless constituent particles are scale invariant. Quantization of these theories typically gives rise to scale-dependent coupling constants that  induce conformal anomalies. Despite the fact that  there are no massless constituent particles, the conformal anomaly has played a major role in cosmology, string theory, the quark-gluon plasma and, more recently,  in modern hydrodynamics. 

In the field of condensed matter, scale invariance and scale anomalies have been discussed  in quantum critical systems,  superfluids, and the unitary Fermi gas in two and three dimensions \cite{Murthy2019,Hu2019,Wang2024}. A recurrent question in this context is whether a quantum anomaly can lead to observable consequences in the macroscopic world. 

The advent of graphene 
\cite{Netal05,Zetal05}, and  Dirac semimetals \cite{AMV18} provides a new platform to study scale invariant systems and to address the above mentioned question.

Dirac semimetals have two important features: Their fermionic quasiparticles behave at low energies as massless Dirac particles, and the hydrodynamic Gurzhi regime is experimentally accessible and, in fact, it has been observed in clean samples both in graphene \cite{CJetal16,BTetal16,BSetal18,PKim2020,ZKetal23,Engdahl2024,Science24,NN25} and in the three dimensional (3D) materials \cite{MKetal16,GMetal18} (for a recent review, see \cite{2023NatR}). In the 3D case, the clean, interacting system is well described by massless Quantum Electrodynamics (QED). The chiral and mixed chiral-gravitational anomalies arising from electromagnetic and gravitational interactions were predicted to induce novel transport phenomena  in Dirac  semimetals \cite{Karl14,Review22,ACH22}. The experimental confirmation of these quantum anomalies \cite{Ong2021,Gooth2017} has been one of the recent highlights of the field. 

The conformal anomaly induced by the renormalization of the electric charge in conformal invariant QED(3+1) was explored in \cite{Chernodub:2016lbo,Chernodub:2017bbd} and new electromagnetic transport  phenomena were predicted to occur in the universe hard to access experimentally. These results 
were adapted recently to 3D Dirac semimetals  where these novel transport phenomena are experimentally accessible \cite{CCV18,ACV19,CV19}. The anomaly associated to the charge renormalization is absent in (2+1) dimensions were the electric charge is not renormalized. But in the Lorentz breaking QED-like models  associated to Dirac matter, there is another running parameter in the Lagrangian: the Fermi velocity \cite{Maria1994,Rev09,IN12,Kost22}. The experimental confirmation of the running of the Fermi velocity in graphene almost twenty years after the theoretical prediction based on QED \cite{Letal08,Geim2011,SPetal11,MEetal12} is a  hallmark of the success of the hybridization between high and low energies brought by the novel materials. 

The  question addressed in this work is what kind of scale anomaly arises from the Fermi velocity renormalization and what are the observable effects associated to it. We will show that there is indeed an anomaly, the trace of the stress tensor being proportional to the Fermi velocity beta function. The renormalization of $v_F$ modifies the hydrodynamic speed of sound in graphene and the anomaly affects the correlation functions -- and observables -- involving $T_\mu^\mu$. In particular, it implies that the bulk viscosity will not be zero in graphene.

\section{QFT model of graphene and renormalization of the Fermi velocity}
\label{sec:grapheneQED}
Dirac semimetals in two and three space dimensions have  Fermi points at or near a two band crossing. The linear dispersion gives rise to an effective low energy continuum  Dirac Hamiltonian
\begin{equation}
{\cal H}=  v_F \int d^n {\bf r} \,\bar{\psi}({\bf r})
\gamma^i\partial_i \psi ({\bf r})\;, 
\label{eq:Ham}
\end{equation}
where  $v_F$ is the Fermi velocity, a parameter related to the properties of the underlying lattice,  and $\gamma^i$ an appropriate set of Dirac matrices\footnote{The Dirac nature of the Hamiltonian is often due to a pseudospin degree of freedom coming from the lattice and is not linked to the spin of the electrons. }. Finally, $n$ labels the number of space dimensions. This Hamiltonian can be promoted to a full QFT action 
\beq
S_0=\int \dd t \, \dd^n {\bm r}\, {\bar \Psi} i \gamma^\mu \partial_\mu \Psi,
\label{freeS}
\eeq
%
with the Dirac matrices redefined as 
\beqn
\gamma^\mu = (\gamma^t, v_F {\bm \gamma})\,.
\label{eq:gamma}
\eeqn
Unless established otherwise, these  will be our Dirac matrices from now on.

In (3+1) dimensions, Coulomb interactions are described by standard QED 
\beq
S=\int \dd t \, \dd^3 {\bm r}\, {\bar \Psi} i \gamma^\mu \nabla_\mu \Psi,
\label{eq:sfull}
\eeq
where $\nabla_\mu$ is the covariant derivative (called Peierls substitution in Condensed Matter)
\beq
\nabla_\mu=\partial_\mu+ieA_\mu.
\label{eq:photon}
\eeq
The  gauge field (photon) has a free Lagrangian  $L_0=F_{\mu\nu}F^{\mu\nu}$ where $F_{\mu\nu}$ is the Maxwell electromagnetic tensor: $F_{\mu\nu}=\partial_\mu A_\nu - \partial_\nu A_\mu$. This is the model for Dirac matter in (3+1) dimensions upon redefining the gamma matrices as in \eqref{eq:gamma}. 

From the structure of the Maxwell kinetic term, we can see that  the wavevector ($k$) dependence of the gauge field propagator  in QFT  is $1/k^2$ in any number of dimensions giving rise to the standard $1/r$ dependence of the Coulomb potential in 3D. The interacting term has the same scale dimension as the free kinetic term and hence the coupling constant is dimensionless. The full model is scale invariant.

The two dimensional QFT case is different.  In planar QED the 2D photon field (Fourier transform of $1/k^2$) depends logarithmically on the distance.   As a consequence, the coupling constant $e$ has  dimension of mass (${\sqrt M}$) and the theory is not scale invariant (it is called super-renomalizable; in fact, there are no ultraviolet infinities in QED(2+1)). There is no scale anomaly associated to the running coupling constant in 2D.
\begin{figure}
\begin{center}
\includegraphics[width=0.6\linewidth]{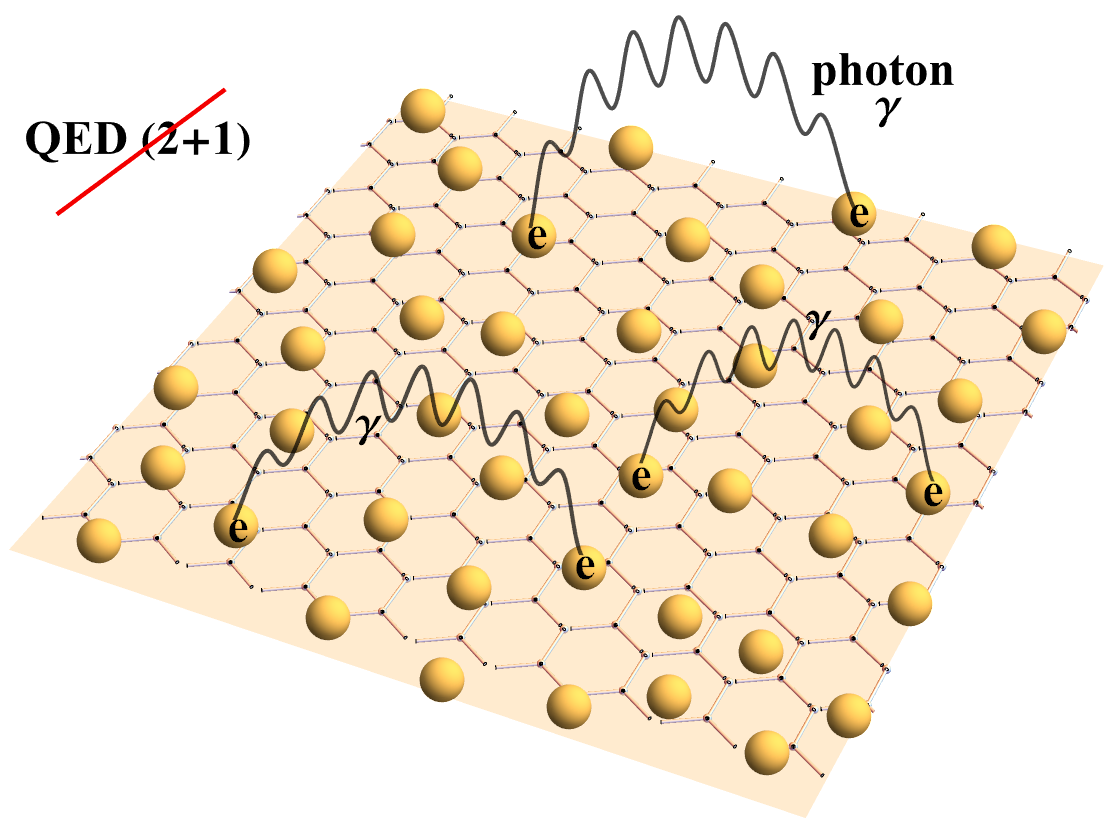}
\caption{Sketch of the Coulomb interaction in graphene. The electrons confined to the plane
can exchange photons in the three spacial dimensions. As a consequence, the interaction between electrons is not captured by QED $(2+1)$.}
\label{fig:brain}
\end{center}
\end{figure}

The case of (2+1) dimensional graphene is more subtle and very interesting. While the free action is given by \eqref{freeS} \cite{W47,SW58,Semenoff84}, the Coulomb interaction is not described by QED(2+1). The reason is that  the charges confined to a plane  can interact exchanging photons in the three dimensions (see Fig.\ref{fig:brain} for a cartoon). Hence we face the problem of coupling the  3D photon (with a $1/k^2$ propagator) to 2D electrons. This problem was already addressed in Ref.~\cite{Maria1994} as follows.
The unscreened Coulomb interaction  considered in solid state physics is 
\begin{equation}
{\cal H}_{int} = \frac{e^2}{2 \pi} \int d^2 {\bf r_1} \int d^2
{\bf r_2} \frac{\bar{\Psi} ( \bf{r_1} ) \Psi ( \bf{r_1} )
\bar{\Psi} (\bf{r_2} ) \Psi ( \bf{r_2} )} {| \bf{r}_1 - \bf{r}_2
|}.
 \label{eq:Hint}
\end{equation}
This static, non--local, non relativistic interaction can be
obtained as a one loop effective action from (\ref{eq:sfull}) by integrating out the vector field with a
$ 1/\vert  k\vert$ propagator and taking the limit   $v_f/c\to 0$. 
This is equivalent to integrating the usual photon propagator over the  component transverse to the plane \cite{Maria1994} and taking the static limit: 
\begin{align}
\Pi_{\mu\nu}&\equiv \langle T A_\mu(t, {\bf r}),A_\nu(t', {\bf r'})\rangle \\ \nonumber
& \approx-i\delta_{\mu\nu}\int \frac{\dd \omega \, \dd^2 {\bm k}}{(2\pi)^3}
\int \frac{\dd k_z}{(2\pi)}
\frac{e^{i{\bm k}\cdot({\bm r}-{\bm r'})}e^{-i\omega(t-t')}}{-\omega^2+{\bm k}^2+k_z^2} 
\\ \nonumber
&=
-i\delta_{\mu\nu}\delta(t-t')\,\frac{1}{2}\int \frac{\dd^2 {\bm k}}{(2\pi)^2}
\frac{1}{\vert {\bm k}\vert}\exp[i{\bm k}\cdot({\bm r}-{\bm r'})].
\end{align}
With this effective  $1/\vert {\bm k}\vert$ propagator, the Coulomb interaction is marginal and the full action \eqref{eq:sfull} is scale invariant. The resulting model was called ``\textit{brain reduced QED}" in \cite{GGM2001} (see also \cite{Marino93}).

Comparing the Hamiltonians \eqref{eq:Ham} and \eqref{eq:Hint}, and using a perturbative scheme, it can be seen that the effective two-dimensional coupling constant  
\beq
\alpha_G=e^2/4\pi v_F
\eeq
is identical to the fine structure constant of QED with the speed of light replaced by the Fermi velocity.

\begin{figure}
\begin{center}
\includegraphics[width=0.7\linewidth]{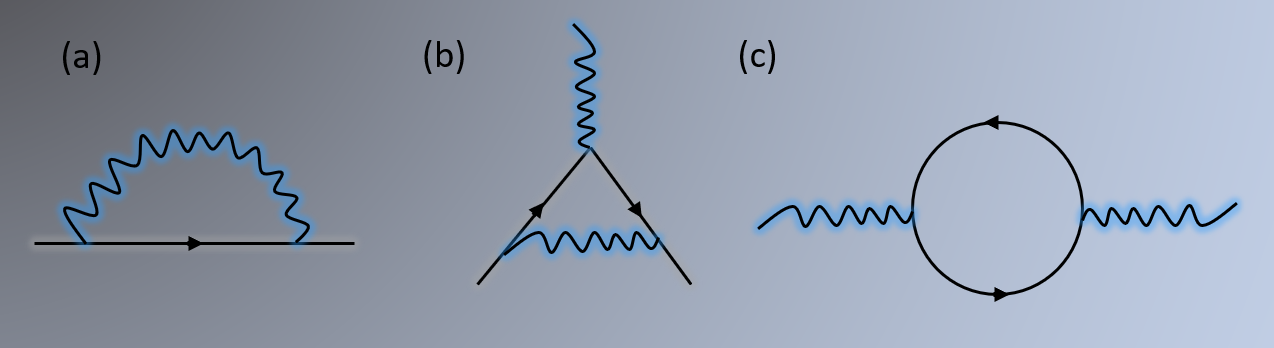}
\caption{Primitively divergent Feynman graphs in QED $(3+1)$. Fermion self--energy (a),  vertex (b) and photon self-energy (c).}
\label{fig:primitive}
\end{center}
\end{figure}

%
%
The renormalization of the graphene model  has been  studied throughly~\cite{Maria1994,GGV99,Maria2011,Rev09}. Of the three one loop graphs shown in Fig. \ref{fig:primitive}, only the electron self--energy diverges and, in the case that we are considering, with a static photon propagator, only the spatial part has a logarithmic divergence. It can be shown to all orders in perturbation theory that the photon propagator is finite and hence the electric charge is not renormalized. All the renormalization of the system comes from the electron propagator. The renormalization of the Fermi velocity and the electron wave function are obtained  from the electron self-energy $\Sigma(\omega,{\bf k})$ computed from  figure \ref{fig:primitive} (a) and from the relations:
\begin{equation}
\Sigma(\omega,{\bf k})=Z_\psi(\omega,{\bf k})
\left[\omega\gamma^0-Z_v(\omega,{\bf k})v \gamma \cdot {\bf k}\right]\;.
\label{zelectron}
\end{equation}

The beta function of the Fermi velocity turns out to be
\beq
\beta_v\equiv \frac{\partial v}{\partial l}=\frac{\partial Z_v}{\partial l}=-\frac{e^2}{16\pi v_F}=-\frac{\alpha_G}{4},
\label{eq:beta:v}
\eeq
where $l=\log\Lambda$ and $\Lambda$ is the ultraviolet cutoff. The 
 Fermi velocity $v_F$ at the energy $\Lambda$ is then given by the renormalization group equation~\cite{Maria1994,JGV10,Maria2011}:
\beqa
v(\Lambda) = v_{F,0} \left[1 - \frac{\alpha_G}{4} \log \frac{\Lambda}{\Lambda_0} \right],
\label{eq:v:omega}
\eeqa
where the initial value $v_{F,0}$ can be taken from experiments. More specifically, at the RG normalization point 
$\Lambda_0 = 125$ MeV, the Fermi velocity equals to~\cite{Geim2007}, $v_F(\Lambda_0) = v_{F,0}= 1.093 \times 10^6$ m/s.

The Fermi velocity renormalization of graphene  was  experimentally verified~\cite{Letal08,Eva09,Geim2011,SPetal11,MEetal12,STM2012}, more than a decade after the first theoretical prediction   \cite{Maria1994}.

\section{Conformal anomaly from the velocity renormalization in graphene}
\label{sec:CA}

The  graphene  action  discussed in the previous section
\beq
S_0=\int \dd t \, \dd^2 {\bm x}\, {\bar \Psi} i \gamma^\mu \nabla_\mu \Psi,
\label{eq:S0}
\eeq
 is invariant under the global scale transformation:
\beqn
{\bm x} \to \lambda {\bm x}, 
\quad
t \to \lambda t, 
\quad
\Psi \to \lambda^{-1} \Psi\,.
\label{eq:scale:transformations}
\eeqn
Notice that these transformations do not depend on the Fermi velocity which is included only into the definition of the gamma functions~\eq{eq:gamma}.

The  transformations~\eq{eq:scale:transformations}  are generated by the dilatation current \cite{Coleman85}:
\beqn
j_D^\mu = T^{\mu\nu} x_\nu\,.
\label{eq:j:D}
\eeqn
On the other hand, the energy-momentum tensor for the action~\eq{eq:S0} is:
\beqn
T^{\mu\nu}  & = & \frac{i}{2} {\bar \Psi} \left(\gamma^\mu \nabla^\nu + \gamma^\nu \nabla^\mu \right) \Psi \nonumber \\
& & 
- i\, \eta^{\mu\nu} {\bar \Psi} \left( \gamma^t \partial_t + v_F \, {\bm \gamma}  \cdot {\bm \nabla}\right) \Psi\,,
\label{eq:Tmunu:psi}
\eeqn 
where, in the second row,  we have made explicit the Fermi velocity presence in the spacial part. The metric tensor
\beqn
\eta^{\mu\nu} = {\mathrm{diag}}\, (+1,-1,-1),
\eeqn
is a flat spacetime metric which does not depend on the Fermi velocity. 
Here, the indices run over both time and space coordinates $\mu, \nu = 0,1,2$ similarly to the scale transformations~\eq{eq:scale:transformations}.
We notice that the energy density operator is as follows:
\beqn
\epsilon \equiv T^{00} = - v_F \, {\bar \Psi} i {\bm \gamma}  \cdot {\bm \nabla} \Psi\,,
\label{eq:epsilon}
\eeqn
where again the Fermi velocity factor is made explicit.

The fermionic energy-momentum tensor~\eq{eq:Tmunu:psi} is given by the variation of the fermionic action $S_0$ with respect to the background metric $g_{\mu\nu}$:
\beqn
T^{\mu\nu} (x) = 2 \frac{\delta S_0[g]}{\delta g_{\mu\nu}(x)} \,,
\qquad
S_0[g] = \int \dd t \, \dd^2 {\bm x} \sqrt{g} \, \cL_\psi(g),
\quad
\label{eq:Tmunu:dS}
\eeqn
with $g = \det (g_{\mu\nu})$, and then calculated in the flat-space limit $g^{\mu\nu}{\to}\, \eta^{\mu\nu}$.

Classically, the dilatation current~\eq{eq:j:D} has zero divergence because the classical equations of motion imply
\beqn
\partial_\mu j_D^\mu = T^{\alpha}_{\ \alpha}\,,
\label{eq:dJ:D}
\eeqn
and the trace of the energy-momentum tensor~\eq{eq:Tmunu:psi} vanishes at the classical level, 
\beqn
T^\alpha_{\ \alpha} = - 2 {\bar \psi} i \gamma^\mu \nabla_\mu \psi \equiv 0,
\eeqn
as a consequence of the classical equation of motion of the theory~\eq{eq:S0},
\beqn
 \gamma^\mu \nabla_\mu \psi = 0.
\eeqn
Therefore, the classical theory is invariant under the scale transformations and the dilatation current~\eq{eq:j:D} vanishes classically: $\partial_\mu j_D^\mu = 0$.

However, scale invariance is broken by quantum fluctuations. Therefore, the quantum expectation value of the right-hand side of Eq.~\eq{eq:dJ:D} is nonzero and, consequently, we expect that on a quantum level the dilatation current~\eq{eq:j:D} is no longer conserved. Below, we will prove this statement and find the expression for the trace of the energy-momentum tensor in the presence of a running Fermi velocity. 

Here, we present a heuristic derivation, that follows a pretty standard procedure for the well-known type of trace anomaly that is associated with charge renormalization in QED.

Consider a small global scale transformation: 
\beqn
\eta_{\mu\nu} \to g_{\mu\nu} = e^{2 \xi} \eta_{\mu\nu}\,,
\qquad
\delta g_{\mu\nu} = 2 \xi \eta_{\mu\nu},\qquad
\label{eq:g:tau}
\eeqn
where the last relation is written for the small scale factor $\xi$ with $|\xi| \ll 1$. 
Equation~\eq{eq:Tmunu:dS}  implies:
\beqn
S \to S_{\xi} = S + \xi \int \dd t \, \dd^2 {\bm x} \,  T^\mu_{\ \mu}(x) + O(\xi^2)\,.
\quad
\label{eq:action:variation}
\eeqn

Thus, the variation of the quantum action with respect to the local dilatation factor gives us the trace of the energy momentum tensor:
\beqn
\frac{\partial S_\psi}{\partial \xi}  = \avr{\int \dd t \, \dd^2 {\bm x}  \, T^\mu_{\ \mu}}.
\label{eq:step}
\eeqn
On the other hand, under the dilatation~\eq{eq:g:tau}, the Fermi velocity changes as follows:
\beqn
v_F \to v_F + \xi \beta_v,
\label{eq:vvariation}
\eeqn
where $\beta_v$ is the beta function associated with the running of the Fermi velocity~\eq{eq:beta:v}.

By combining Eqs.~\eq{eq:action:variation} and \eq{eq:vvariation}, we obtain
\beqn
\frac{\partial S_\psi}{\partial \xi} = \beta_v \int \dd t \, \dd^2 {\bm x} \, \avr{\bar \Psi i {\bs \gamma} {\bs D} \Psi},
\label{eq:variation:S}
\eeqn
that implies -- via Eq.~\eq{eq:step} -- the following equation for the conformal anomaly:
\beqn
\avr{T^\mu_{\ \mu}} = \beta_v  \avr{\bar \Psi i {\bs \gamma} {\bs \nabla} \Psi}.
\label{eq:new:1}
\eeqn
This is the main result of this work.

Repeating the derivation with a coordinate-dependent factor $\xi = \xi(x)$, using a variation instead of the differential operator in Eq.~\eq{eq:variation:S}, we also get a relation for a matrix element with arbitrary insertions of $T^\mu_\mu$'s:
\begin{equation}
\avr{T^\mu_{\ \mu}(x_1) \dots T^\mu_{\ \mu}(x_n) } =
\beta_v^n \avr{\bar \Psi(x_1) i {\bs \gamma} {\bs \nabla} \Psi(x_1) \dots \bar \Psi(x_n) i {\bs \gamma} {\bs \nabla} \Psi(x_n)}.
\label{eq:new:2}
\end{equation}
This formula implies the validity of the local relation:
\beqn
T^\mu_{\ \mu}(x) = \beta_v \bar \Psi(x) i {\bs \gamma} {\bs \nabla} \Psi(x).
\label{eq:new:3}
\eeqn
We notice that this relation is valid only for the matrix elements (expectation values) and it should not be considered as an operator relation. 

Exactly the same situation holds in 3+1 dimensional QED, where the trace of the energy-momentum tensor is {\it{locally}} associated with a (gauge) field with a beta-function as a pre-factor.


\section{Physical consequences of the anomaly}

\subsection{Thermodynamics}

The standard anomaly in QED, associated with the renormalization of the electric charge, reveals itself in the classical conformal-violating background field \cite{Chernodub16}. Similarly, the conformal anomaly will reveal itself at finite density and/or finite temperature, where the right-hand-side of Eq.~\eq{eq:new:1} becomes nonzero. 
The right-hand side of Eq.~\eq{eq:new:1} is, up to a proportionality coefficient, the energy density of the Dirac gas~\eq{eq:epsilon}:
\beqn
E \equiv \avr{T^{00}} = - v_F \avr{\bar \Psi i {\bs \gamma} {\bs \nabla} \Psi}.
\label{eq:E}
\eeqn
Therefore, the conformal anomaly equation~\eq{eq:new:1} may also be represented in the form:
\beqn
\avr{T^\mu_{\ \mu}} = - \frac{\beta_v}{v_F} E.
\eeqn
Taking into account that the trace of the energy-momentum tensor is related to the pressure $P$ and energy $E$ by, $\avr{(T_\psi)^\mu_{\ \mu}} = E - 2 P$, we get:
\beqn
\left(1 + \frac{\beta_v}{v_F}\right) E = 2 P.
\label{eq:new:2}
\eeqn
Thus, the presence of the beta function $\beta_v$, associated with the renormalization of the Fermi velocity,  modifies the conformal equation of state $E = 2 P$.
As a consequence, the density of state and all the thermodynamic quantities get also modified. In particular, the electronic specific heat
of graphene, a quantity that is now accurately measured \cite{Ce21} will suffer a logarithmic suppression most noticeable at intermediate to high temperatures (this suppression was noticed in the early work \cite{Vafek2007}).

\subsection{Speed of sound}

The barotropic equation of state of a perfect fluid or a relativistic gas in special relativity reads:
\beqn
\frac{\partial P}{\partial E} = \frac{v_s^2}{c^2},
\eeqn
where $c_s$ is the speed of sound and $c$ is the speed of light (the latter is played by the Fermi velocity in graphene, $c \leftrightarrow  v_F$). 

Due to the modification of the equation of state by the conformal anomaly~\eq{eq:new:2}, the speed of hydrodynamical sound gets modified:
\beq
v_s(T) = \frac{v_F(T)}{\sqrt{2}} {\left(1 - \frac{\beta_v}{v_F(T)}\right)}^{-1/2}
\label{eq:vs}
\eeq
where temperature $T$ plays of the energy scale $\Lambda$. 
The speed of sound of a free relativistic  Dirac fluid in d dimensions obeys the rule $v_s<c/\sqrt{d}$.
The renormalization of the Fermi velocity does not lead to the breaking of the ``relativistic sound barrier'' in Eq.~\eq{eq:vs} because $\beta_v \leqslant 0$ and
\beqn
v_s(T) \leqslant \frac{v_F(T)}{\sqrt{2}}\,.
\eeqn
The sound velocity can be accurately measured by terahertz photocurrent
nanoscopy\cite{Pricipi23}.
\subsection{Bulk viscosity}

The bulk viscosity (or sometimes ``second viscosity'') $\zeta$ characterizes how a system of particles deviates from thermodynamic equilibrium in the course of uniform expansion (a nice introduction can be found in Ref.~\cite{Sharma2023}). This parameter, assumed to be very small or zero in many classical fluids, has proven to be relevant in condensed matter systems  \cite{Enss2019,FUJII2023}. A finite bulk viscosity is necessary to eliminate the well-known discrepancy between theory and observation of sound absorption in liquids at very high frequencies \cite{PhysRev.75.1415}. 
In scale-invariant theories, the bulk viscosity is identically zero since, upon a uniform expansion, an equilibrium distribution function remains in equilibrium (\textit{albeit} at an adapted temperature) if the dispersion relation exhibits scale invariance. More technically, the Ward identity associated to scale invariance implies that $\zeta$ identically vanishes at any temperature and frequency (see \textit{e.g.} appendix B in \cite{ENSS2011770} or \cite{weinberg1972gravitation}).

In a rotationally invariant two-dimensional system, the shear $\eta$ and the bulk $\zeta$ viscosities are related to the spatial components $i,j,l,m = 1,2$ of the correlation function of the energy-momentum tensor $T_{ij}(x)$ via Kubo's formula \cite{PhysRevB.86.245309},
\beqn
&& \eta(\omega)\left( \delta_{il}\delta_{jm}+\delta_{im}\delta_{jl}-\delta_{ij}\delta_{lm}\right)
+\zeta(\omega)\delta_{ij}\delta_{lm}
\label{eq:viscosity:def}\\
&&=\frac{1}{\omega}\lim_{{\bs k} \to 0} \int d^2 x
\int_0^\infty dt\, e^{i(\omega t- {\bs k} {\bs x})} \avr{[T_{ij}(t, {\bs x}),T_{lm}(0, {\bs 0})]},
\quad \nonumber
\eeqn
where the vanishing of the two-dimensional wavevector~$\bs k$ corresponds to the uniform limit with the frequency $\omega$ kept finite.

The static limit $\omega \to 0$ of the correlation function~\eq{eq:viscosity:def} with contracted $i,j$ and $l,m$ indices gives us the bulk viscosity~$\zeta$ expressed via the spatial trace of the energy-momentum tensor:
\beqn
\zeta = \lim_{\omega\to 0} \frac{1}{4\omega} \int\nolimits_0^\infty dt \int d^2 x \, e^{i\omega t} \avr{ [T_{ii}(x),T_{jj}(0)]},
\label{eq:viscosity:0}
\eeqn
where we also took the limit ${\bs k} \to 0$ explicitly, assumed the sum over the silent indices, and with $x = (t, {\bs x})$.
Expression \eqref{eq:new:3} implies that the bulk viscosity is finite in the graphene fluid and proportional to the square of the beta function for the Fermi velocity.

In the relaxation time approximation,  the bulk viscosity in d spacial dimensions is directly linked to the shear viscosity $\eta$ as \cite{PhysRevC.85.044909}
\begin{equation}
    \zeta = \# \eta \left( \frac{1}{d} - \frac{v_s^2}{c^2} \right)^2	, \label{lala}
\end{equation}
as first recognized by Weinberg for a photon gas coupled to matter \cite{1971ApJ...168..175W} (where $\# \approx 15$). 
In the graphene case $c=v_F$.

Eq. \eqref{lala} is reminiscent of an analogous expression derived from AdS-CFT and valid in the strong coupling limit \cite{BUCHEL2008286}, 
\begin{equation}
    \zeta > 2 \eta \left( \frac{1}{d} - \frac{c_s^2}{c^2} \right).
\end{equation}

The bulk viscosity can be measured using several experimental techniques including Brillouin spectroscopy, Laser transient grating spectroscopy, and acoustic spectroscopy \cite{10.1063/1.3095471}.

\section{Discussion and future directions}
\label{sec:end}
The  results  on the conformal anomaly induced by the renormalization of the Fermi velocity in graphene,  can be easily extended to the three dimensional Dirac and Weyl semimetals \cite{AMV17}.
Their low energy dynamics is described by a Lorentz breaking QED(3+1) \eqref{eq:sfull} with the inclusion of the Fermi velocity in the spacial Dirac matrices given in \eqref{eq:gamma}. 
In this model, the divergence of the photon polarization (diagram in Fig. \ref{fig:primitive}c) gives rise to the renormalization of the electric charge what, in turn, induces the standard QED conformal anomaly \cite{Chernodub:2016lbo,Chernodub:2017bbd} mentioned earlier: 
\beq
<T^\mu_\mu(x)> = \frac{\beta_e}{2e} F^{\mu\nu}(x) F_{\mu\nu}(x).
\eeq
This anomaly has already been studied in the condensed matter domain and it has been shown to give rise to new transport phenomena experimentally accessible \cite{CCV18,ACV19,CV19}. 

But in these 3D materials the Fermi velocity is also renormalized \cite{IN12,RJH16,PFV18,Kost22}, their beta function is also negative and proportional to the coupling constant. The analysis done in this work, in particular eq. \eqref{eq:new:1},
goes through straightforwardly, improving the chances for the experimental observation of the phenomenon.

In addition to  the finite bulk viscosity, the conformal anomaly modifies the equation of state of and affects the thermodynamic and hydrodynamic properties giving rise  to other observable quantities experimentally accessible. In particular, the specific heat ..

The corrections discussed in this work   are proportional to the beta function of the Fermi velocity. In graphene as well as in the 3D Dirac semimetals the beta function is proportional to the fine structure constant of the materials 
$\alpha_v=e^2/4\pi\epsilon v_F$ (we have introduced the dielectric constant of the environment, an important parameter which is $\epsilon=1$ in vacuum but can be tuned in the material experiments. ) 
It is important to notice that, due to the relation  $v_F/c\sim 10^{-2}$, the fine structure constant both in graphene and in 3D Dirac semimetals is not a small parameter as in QED  and the corrections need not be small.
The 3D materials are affected by multiple factors (they usually are slightly doped, have various types of disorder, other trivial bands cross the Fermi surface), but graphene can be extremely clean and its coupling constant in vacuum is already large, \textit{i.e.}, $\alpha_G\sim 2$.
The  fact that graphene does not show strong coupling behavior is one of the most interesting aspects still remaining in the field despite some attempts of an explanation \cite{Fradkin2010a,Fradkin2010b,Sarma14,Herbut2018}.
The experiments associated to the scale anomaly will provide direct access to the value of $\alpha$ at the scale of the given experimental setup. 


\section*{Acknowledgements}
M.B. acknowledges the support of the Shanghai Municipal Science and Technology Major Project (Grant No.2019SHZDZX01) and the sponsorship from the Yangyang Development Fund.
The work of M.N.C. has been partially supported by the French National Agency for Research (ANR) within the project PROCURPHY ANR-23-CE30-0051-02.

\bibliographystyle{JHEP}
\bibliography{Graphene}

\providecommand{\href}[2]{#2}\begingroup\raggedright\begin{thebibliography}{10}

\bibitem{Nakayama:2013is}
Y.~Nakayama, \emph{{Scale invariance vs conformal invariance}},
  \href{http://dx.doi.org/10.1016/j.physrep.2014.12.003}{\emph{Phys. Rept.}
  {\bf 569} (2015) 1--93}, [\href{http://arxiv.org/abs/1302.0884}{{\tt
  1302.0884}}].

\bibitem{10.1093/acprof:oso/9780199577224.001.0001}
H.~Nishimori and G.~Ortiz, \emph{Elements of Phase Transitions and Critical
  Phenomena}.
\newblock Oxford University Press, 2010.

\bibitem{doi:10.1098/rsif.2017.0662}
Y.~Khaluf, E.~Ferrante, P.~Simoens and C.~Huepe, \emph{Scale invariance in
  natural and artificial collective systems: a review}, {\emph{Journal of The
  Royal Society Interface} {\bf 14} (2017) 20170662}.

\bibitem{Zinn-Justin:2010}
J.~Zinn-Justin, \emph{{C}ritical {P}henomena: field theoretical approach},
  \href{http://dx.doi.org/10.4249/scholarpedia.8346}{\emph{Scholarpedia} {\bf
  5} (2010) 8346}.

\bibitem{Murthy2019}
P.~A. Murthy, N.~Defenu, L.~Bayha, M.~Holten, P.~M. Preiss, T.~Enss et~al.,
  \emph{Quantum scale anomaly and spatial coherence in a 2d fermi superfluid},
  \href{http://dx.doi.org/10.1126/science.aau4402}{\emph{Science} {\bf 365}
  (2019) 268--272}.

\bibitem{Hu2019}
H.~Hu, B.~C. Mulkerin, U.~Toniolo, L.~He and X.-J. Liu, \emph{Reduced quantum
  anomaly in a quasi-two-dimensional fermi superfluid: Significance of the
  confinement-induced effective range of interactions},
  \href{http://dx.doi.org/10.1103/PhysRevLett.122.070401}{\emph{Phys. Rev.
  Lett.} {\bf 122} (Feb, 2019) 070401}.

\bibitem{Wang2024}
L.~Wang, X.~Yan, J.~Min, D.~Sun, X.~Xie, S.-G. Peng et~al., \emph{Scale
  invariance of a spherical unitary fermi gas},
  \href{http://dx.doi.org/10.1103/PhysRevLett.132.243403}{\emph{Phys. Rev.
  Lett.} {\bf 132} (Jun, 2024) 243403}.

\bibitem{Netal05}
K.~S. Novoselov, A.~K. Geim, S.~V. Morozov, D.~Jiang, M.~I. Katsnelson, I.~V.
  Grigorieva et~al., \emph{Two-dimensional gas of massless dirac fermions in
  graphene}, {\emph{Nature} {\bf 438} (2005) 197}.

\bibitem{Zetal05}
Y.~Zhang, Y.-W. Tan, H.~L. Stormer and P.~Kim, \emph{Experimental observation
  of the quantum hall effect and berry phase in graphene}, {\emph{Nature} {\bf
  438} (2005) 201}.

\bibitem{AMV18}
N.~P. Armitage, E.~J. Mele and A.~Vishwanath, \emph{Weyl and dirac semimetals
  in three-dimensional solids},
  \href{http://dx.doi.org/10.1103/RevModPhys.90.015001}{\emph{Rev. Mod. Phys.}
  {\bf 90} (Jan, 2018) 015001}.

\bibitem{CJetal16}
J.~Crossno, J.~K. Shi, K.~Wang, X.~Liu, A.~Harzheim, A.~Lucas et~al.,
  \emph{Observation of the dirac fluid and the breakdown of the wiedemann-franz
  law in graphene}, {\emph{Science} {\bf 351} (2016) 1058--1061}.

\bibitem{BTetal16}
D.~A. Bandurin, I.~Torre, R.~K. Kumar, M.~Ben~Shalom, A.~Tomadin, A.~Principi
  et~al., \emph{Negative local resistance caused by viscous electron backflow
  in graphene}, {\emph{Science} {\bf 351} (2016) 1055--1058}.

\bibitem{BSetal18}
D.~A. Bandurin, A.~V. Shytov, L.~S. Levitov, R.~K. Kumar, A.~I. Berdyugin,
  M.~Ben~Shalom et~al., \emph{Fluidity onset in graphene}, {\emph{Nature
  Communications} {\bf 9} (2018) 4533}.

\bibitem{PKim2020}
M.~J.~H. Ku, T.~X. Zhou, Q.~Li, Y.~J. Shin, J.~K. Shi, C.~Burch et~al.,
  \emph{Imaging viscous flow of the dirac fluid in graphene}, {\emph{Nature}
  {\bf 583} (2020) 537--541}.

\bibitem{ZKetal23}
Z.~J. Krebs, W.~A. Behn, S.~Li, K.~J. Smith, K.~Watanabe, T.~Taniguchi et~al.,
  \emph{Imaging the breaking of electrostatic dams in graphene for ballistic
  and viscous fluids}, {\emph{Science} {\bf 379} (2023) 671--676}.

\bibitem{Engdahl2024}
J.~N. Engdahl, A.~C. Keser, T.~Schmidt and O.~P. Sushkov, \emph{Driving viscous
  hydrodynamics in bulk electron flow in graphene using micromagnets},
  {\emph{Phys. Rev. B} (2024) }, [\href{http://arxiv.org/abs/2312.04896}{{\tt
  2312.04896}}].

\bibitem{Science24}
M.~L. Palm, C.~Ding, W.~S. Huxter, T.~Taniguchi, K.~Watanabe and C.~L. Degen,
  \emph{Observation of current whirlpools in graphene at room temperature},
  \href{http://dx.doi.org/10.1126/science.adj2167}{\emph{Science} {\bf 384}
  (2024) 465--469}.

\bibitem{NN25}
M.~Kravtsov, A.~L. Shilov, Y.~Yang, T.~Pryadilin, M.~A. Kashchenko, O.~Popova
  et~al., \emph{Viscous terahertz photoconductivity of hydrodynamic electrons
  in graphene}, {\emph{Nature Nanotechnology} {\bf 20} (2025) 51--56}.

\bibitem{MKetal16}
P.~J.~W. Moll, P.~Kushwaha, N.~Nandi, B.~Schmidt and A.~P. Mackenzie,
  \emph{Evidence for hydrodynamic electron flow in {$PdCoO_2$}},
  {\emph{Science} {\bf 351} (2016) 1061--1064}.

\bibitem{GMetal18}
J.~Gooth, F.~Menges, N.~Kumar, V.~S{\"u}?, C.~Shekhar, Y.~Sun et~al.,
  \emph{Thermal and electrical signatures of a hydrodynamic electron fluid in
  tungsten diphosphide}, {\emph{Nature Communications} {\bf 9} (2018) 4093}.

\bibitem{2023NatR}
G.~{Varnavides}, A.~{Yacoby}, C.~{Felser} and P.~{Narang}, \emph{{Charge
  transport and hydrodynamics in materials}},
  \href{http://dx.doi.org/10.1038/s41578-023-00597-3}{\emph{Nature Reviews
  Materials} {\bf 8} (Nov., 2023) 726--741}.

\bibitem{Karl14}
K.~Landsteiner, \emph{Anomalous transport of weyl fermions in weyl semimetals},
  {\emph{Phys. Rev. B} {\bf 89} (2014) 075124}.

\bibitem{Review22}
M.~N. Chernodub, Y.~Ferreiros, A.~G. Grushin, K.~Landsteiner and M.~A.
  Vozmediano, \emph{Thermal transport, geometry, and anomalies},
  \href{http://dx.doi.org/https://doi.org/10.1016/j.physrep.2022.06.002}{\emph{Physics
  Reports} {\bf 977} (2022) 1--58}.

\bibitem{ACH22}
R.~Arouca, A.~Cappelli and T.~H. Hansson, \emph{{Quantum Field Theory Anomalies
  in Condensed Matter Physics}}, {\emph{SciPost Phys. Lect. Notes} (2022) 62}.

\bibitem{Ong2021}
N.~P. Ong and S.~Liang, \emph{Experimental signatures of the chiral anomaly in
  dirac and weyl semimetals},
  \href{http://dx.doi.org/10.1038/s42254-021-00310-9}{\emph{Nature Reviews
  Physics} {\bf 3} (May, 2021) 394–404}.

\bibitem{Gooth2017}
J.~Gooth et~al., \emph{Experimental signatures of the mixed axial gravitational
  anomaly in the weyl semimetal {NbP}}, {\emph{Nature} {\bf 547} (2017) 23005}.

\bibitem{Chernodub:2016lbo}
M.~N. Chernodub, \emph{Anomalous transport due to the conformal anomaly},
  {\emph{Phys. Rev. Lett.} {\bf 117} (2016) 141601}.

\bibitem{Chernodub:2017bbd}
M.~N. Chernodub and M.~A. Zubkov, \emph{Scale magnetic effect in quantum
  electrodynamics and the wigner-weyl formalism}, {\emph{Phys. Rev. D} {\bf 96}
  (2017) 056006}.

\bibitem{CCV18}
M.~N. Chernodub, A.~Cortijo and M.~A.~H. Vozmediano, \emph{Generation of a
  {Nernst} current from the conformal anomaly in {Dirac} and {Weyl}
  semimetals}, {\emph{Phys. Rev. Lett.} {\bf 120} (2018) 206601}.

\bibitem{ACV19}
V.~Arjona, M.~N. Chernodub and M.~A.~H. Vozmediano, \emph{Fingerprints of the
  conformal anomaly on the thermoelectric transport in {Dirac} and {Weyl}
  semimetals: Result from a {Kubo} formula}, {\emph{Phys. Rev. B} {\bf 99}
  (2019) 235123}.

\bibitem{CV19}
M.~N. Chernodub and M.~A.~H. Vozmediano, \emph{Direct measurement of a beta
  function and an indirect check of the {Schwinger} effect near the boundary in
  {Dirac} and {Weyl} semimetals}, {\emph{Phys. Rev. Research} {\bf 1} (2019)
  032002}.

\bibitem{Maria1994}
J.~Gonz{\'a}lez, F.~Guinea and M.~Vozmediano, \emph{Non-fermi liquid behavior
  of electrons in the half-filled honeycomb lattice (a renormalization group
  approach)}, {\emph{Nuclear Physics B} {\bf 424} (1994) 595--618}.

\bibitem{Rev09}
A.~H. Castro~Neto, F.~Guinea, N.~M.~R. Peres, K.~S. Novoselov and A.~K. Geim,
  \emph{The electronic properties of graphene},
  \href{http://dx.doi.org/10.1103/RevModPhys.81.109}{\emph{Rev. Mod. Phys.}
  {\bf 81} (Jan, 2009) 109--162}.

\bibitem{IN12}
H.~Isobe and N.~Nagaosa, \emph{Theory of a quantum critical phenomenon in a
  topological insulator: (3+1)-dimensional quantum electrodynamics in solids},
  {\emph{Phys. Rev. B} {\bf 86} (2012) 165127}.

\bibitem{Kost22}
V.~A. Kosteleck\'y, R.~Lehnert, N.~McGinnis, M.~Schreck and B.~Seradjeh,
  \emph{Lorentz violation in dirac and weyl semimetals}, {\emph{Phys. Rev.
  Res.} {\bf 4} (2022) 023106}.

\bibitem{Letal08}
Z.~Q. Li, E.~A. Henriksenand, Z.~Jiang, Z.~Ha, M.~C. Martin, P.~Kim et~al.,
  \emph{Dirac charge dynamics in graphene by infrared spectroscopy},
  {\emph{Nature Phys.} {\bf 4} (2008) 532}.

\bibitem{Geim2011}
D.~C. Elias, R.~V. Gorbachev, A.~S. Mayorov, S.~V. Morozov, A.~A. Zhukov,
  P.~Blake et~al., \emph{Dirac cones reshaped by interaction effects in
  suspended graphene}, \href{http://dx.doi.org/10.1038/nphys2049}{\emph{Nature
  Physics} {\bf 7} (2011) 701--704}.

\bibitem{SPetal11}
D.~A. Siegel, C.~Park, C.~Hwang, J.Deslippe, A.~V. Fedorov, S.~G. Louie et~al.,
  \emph{Many-body interactions in quasi-freestanding graphene}, {\emph{PNAS}
  {\bf 108} (2011) 11365}.

\bibitem{MEetal12}
A.~S. Mayorov, D.~C. Elias, I.~S. Mukhin, S.~V. Morozov, L.~A. Ponomarenko,
  K.~S. Novoselov et~al., \emph{How close can one approach the dirac point in
  graphene experimentally?}, {\emph{Nano Letters} {\bf 12} (2012) 4629--4634}.

\bibitem{W47}
P.~R. Wallace, \emph{The band theory of graphite}, {\emph{Phys. Rev.} {\bf 71}
  (1947) 622}.

\bibitem{SW58}
J.~C. Slonczewski and P.~R. Weiss, \emph{Band structure of graphite},
  {\emph{Phys. Rev.} {\bf 109} (1958) 272}.

\bibitem{Semenoff84}
G.~W. Semenoff, \emph{Condensed-matter simulation of a three-dimensional
  anomaly}, \href{http://dx.doi.org/10.1103/PhysRevLett.53.2449}{\emph{Phys.
  Rev. Lett.} {\bf 53} (Dec, 1984) 2449--2452}.

\bibitem{GGM2001}
E.~V. Gorbar, V.~P. Gusynin and V.~A. Miransky, \emph{Dynamical chiral symmetry
  breaking on a brane in reduced {QED}}, {\emph{Phys. Rev. D} {\bf 64} (2001)
  105028}.

\bibitem{Marino93}
E.~Marino, \emph{Quantum electrodynamics of particles on a plane and the
  chern-simons theory}, {\emph{Nuclear Physics B} {\bf 408} (1993) 551--564}.

\bibitem{GGV99}
J.~Gonz\'alez, F.~Guinea and M.~A.~H. Vozmediano, \emph{Marginal-fermi-liquid
  behavior from two-dimensional coulomb interaction}, {\emph{Phys. Rev. B} {\bf
  59} (1999) R2474--R2477}.

\bibitem{Maria2011}
Vozmediano, \emph{Renormalization group aspects of graphene},
  {\emph{Philosophical Transactions of the Royal Society A} {\bf 369} (2011)
  2625}.

\bibitem{JGV10}
F.~de~Juan, A.~G. Grushin and M.~Vozmediano, \emph{Renormalization of coulomb
  interactions in graphene: Determining observable quantities}, {\emph{Phys.
  Rev. B} {\bf 82} (2010) 125409}.

\bibitem{Geim2007}
R.~S. Deacon, K.-C. Chuang, R.~J. Nicholas, K.~S. Novoselov and A.~K. Geim,
  \emph{Cyclotron resonance study of the electron and hole velocity in graphene
  monolayers}, {\emph{Phys. Rev. B} {\bf 76} (Aug, 2007) 081406}.

\bibitem{Eva09}
G.~Li, A.~Luican and E.~Y. Andrei, \emph{Scanning tunneling spectroscopy of
  graphene on graphite}, {\emph{Phys. Rev. Lett.} {\bf 102} (2009) 176804}.

\bibitem{STM2012}
J.~Chae, S.~Jung, A.~F. Young, C.~R. Dean, L.~Wang, Y.~Gao et~al.,
  \emph{Renormalization of the graphene dispersion velocity determined from
  scanning tunneling spectroscopy}, {\emph{Phys. Rev. Lett.} {\bf 109} (2012)
  116802}.

\bibitem{Coleman85}
S.~Coleman, \emph{Aspects of Symmetry: Selected {Erice} Lectures}.
\newblock Cambridge University Press, 1985,
  \href{http://dx.doi.org/10.1017/CBO9780511565045}{10.1017/CBO9780511565045}.

\bibitem{Chernodub16}
M.~N. Chernodub, \emph{Anomalous transport due to the conformal anomaly},
  {\emph{Phys. Rev. Lett.} {\bf 117} (2016) 141601}.

\bibitem{Ce21}
M.~A. Aamir, J.~N. Moore, X.~Lu, P.~Seifert, D.~Englund, K.~C. Fong et~al.,
  \emph{Ultrasensitive calorimetric measurements of the electronic heat
  capacity of graphene}, {\emph{Nano Letters} (2021) 5330--5337}.

\bibitem{Vafek2007}
O.~Vafek, \emph{Anomalous thermodynamics of coulomb-interacting massless dirac
  fermions in two spatial dimensions},
  \href{http://dx.doi.org/10.1103/PhysRevLett.98.216401}{\emph{Phys. Rev.
  Lett.} {\bf 98} (May, 2007) 216401}.

\bibitem{Pricipi23}
D.~B. Ruiz et~al., \emph{Experimental signatures of the transition from
  acoustic plasmon to electronic sound in graphene}, {\emph{Science Advances}
  {\bf 9} (2023) eadi0415}.

\bibitem{Sharma2023}
B.~Sharma, R.~Kumar and S.~Pareek, \emph{Bulk viscosity of dilute gases and
  their mixtures}, {\emph{Fluids} {\bf 8} (2023) }.

\bibitem{Enss2019}
T.~Enss, \emph{Bulk viscosity and contact correlations in attractive fermi
  gases}, \href{http://dx.doi.org/10.1103/PhysRevLett.123.205301}{\emph{Phys.
  Rev. Lett.} {\bf 123} (2019) 205301}.

\bibitem{FUJII2023}
K.~Fujii and T.~Enss, \emph{Bulk viscosity of resonantly interacting fermions
  in the quantum virial expansion}, {\emph{Annals of Physics} {\bf 453} (2023)
  169296}.

\bibitem{PhysRev.75.1415}
L.~N. Liebermann, \emph{The second viscosity of liquids},
  \href{http://dx.doi.org/10.1103/PhysRev.75.1415}{\emph{Phys. Rev.} {\bf 75}
  (May, 1949) 1415--1422}.

\bibitem{ENSS2011770}
T.~Enss, R.~Haussmann and W.~Zwerger, \emph{Viscosity and scale invariance in
  the unitary fermi gas},
  \href{http://dx.doi.org/https://doi.org/10.1016/j.aop.2010.10.002}{\emph{Annals
  of Physics} {\bf 326} (2011) 770--796}.

\bibitem{weinberg1972gravitation}
S.~Weinberg, \emph{Gravitation and Cosmology: Principles and Applications of
  the General Theory of Relativity}.
\newblock Wiley, 1972.

\bibitem{PhysRevB.86.245309}
B.~Bradlyn, M.~Goldstein and N.~Read, \emph{Kubo formulas for viscosity: Hall
  viscosity, ward identities, and the relation with conductivity},
  \href{http://dx.doi.org/10.1103/PhysRevB.86.245309}{\emph{Phys. Rev. B} {\bf
  86} (Dec, 2012) 245309}.

\bibitem{PhysRevC.85.044909}
K.~Dusling and T.~Sch\"afer, \emph{Bulk viscosity, particle spectra, and flow
  in heavy-ion collisions},
  \href{http://dx.doi.org/10.1103/PhysRevC.85.044909}{\emph{Phys. Rev. C} {\bf
  85} (Apr, 2012) 044909}.

\bibitem{1971ApJ...168..175W}
S.~{Weinberg}, \emph{{Entropy Generation and the Survival of Protogalaxies in
  an Expanding Universe}},
  \href{http://dx.doi.org/10.1086/151073}{\emph{Astrophysical Journal} {\bf
  168} (Sept., 1971) 175}.

\bibitem{BUCHEL2008286}
A.~Buchel, \emph{Bulk viscosity of gauge theory plasma at strong coupling},
  \href{http://dx.doi.org/https://doi.org/10.1016/j.physletb.2008.03.069}{\emph{Physics
  Letters B} {\bf 663} (2008) 286--289}.

\bibitem{10.1063/1.3095471}
A.~S. Dukhin and P.~J. Goetz, \emph{Bulk viscosity and compressibility
  measurement using acoustic spectroscopy},
  \href{http://dx.doi.org/10.1063/1.3095471}{\emph{The Journal of Chemical
  Physics} {\bf 130} (03, 2009) 124519}.

\bibitem{AMV17}
N.~Armitage, E.~J. Mele and A.~Vishwanath, \emph{Weyl and dirac semimetals in
  three dimensional solids}, {\emph{Rev. Mod. Phys.} {\bf 90} (2018) 015001}.

\bibitem{RJH16}
B.~Roy, V.~Juri{\v{c}}i{\'{c}} and I.~F. Herbut, \emph{Emergent lorentz
  symmetry near fermionic quantum critical points in two and three dimensions},
  {\emph{Journal of High Energy Physics} {\bf 4} (2016) 018}.

\bibitem{PFV18}
O.~Pozo, Y.~Ferreiros and M.~A.~H. Vozmediano, \emph{Anisotropic fixed points
  in dirac and weyl semimetals}, {\emph{Phys. Rev. B} {\bf 98} (2018) 115122}.

\bibitem{Fradkin2010a}
P.~{Abbamonte}, J.~P. {Reed}, B.~{Uchoa}, Y.~I. {Joe}, E.~{Fradkin} and
  D.~{Casa}, \emph{{Why isn't graphene a strongly correlated electron
  system?}},  in \emph{APS March Meeting Abstracts}, vol.~2010 of \emph{APS
  Meeting Abstracts}, p.~B21.001, Mar., 2010.

\bibitem{Fradkin2010b}
J.~P. Reed, B.~Uchoa, Y.~I. Joe, Y.~Gan, D.~Casa, E.~Fradkin et~al., \emph{The
  effective fine-structure constant of freestanding graphene measured in
  graphite}, {\emph{Science} {\bf 330} (2010) 805--808}.

\bibitem{Sarma14}
J.~Hofmann, E.~Barnes and S.~Das~Sarma, \emph{Why does graphene behave as a
  weakly interacting system?}, {\emph{Phys. Rev. Lett.} {\bf 113} (2014)
  105502}.

\bibitem{Herbut2018}
H.-K. Tang, J.~N. Leaw, J.~N.~B. Rodrigues, I.~F. Herbut, P.~Sengupta, F.~F.
  Assaad et~al., \emph{The role of electron-electron interactions in
  two-dimensional dirac fermions}, {\emph{Science} {\bf 361} (2018) 570--574}.

\end{thebibliography}\endgroup

\end{document}